\documentclass[twocolumn]{revtex4}
\usepackage[utf8]{inputenc}
\usepackage{graphicx}

\begin{document}

\title{Topological charge, helicity and vorticity conservation and the reverse spin-current model in the II-nd type multifferoics}

\author{Pavel A. Andreev}
\email{andreevpa@my.msu.ru}
\affiliation{Department of General Physics, Faculty of physics, Lomonosov Moscow State University, Moscow, Russian Federation, 119991.}

\date{\today}

\begin{abstract}
The topological charge and its density are related to the spin vorticity,
while the spin vorticity is a part of full vorticity of the medium.
The full vorticity is related to an integral of motion called the hydrodynamic helicity.
Relation between two integrals of motion (the topological charge and the hydrodynamic helicity) is traced.
The role of the spin-current model of the electric polarization formation due to the specific spin distribution in the helicity conservation is demonstrated.
The reverse spin-current model is suggested to demonstrate the contribution of the polarization in the spin evolution equation,
its importance for the helicity conservation is shown as well.
Spin-field model of the electric polarization is the second model for the electric polarization
(the deformation, firstly)
formation due to the specific spin distribution,
which appears from the same principles,
but it has nonstationary origin.
Moreover,
the reverse spin-current model
gives a mechanism for the spin structure formation due to the electric polarization in the system.
It also appears as the requirement for the conservation of the helicity and the topological charge.
\end{abstract}


\maketitle


\section{Introduction}

Integrals of motion are the most essential of all parameters describing any physical system.
Conservation of these parameters allows to use them for the description of the evolution of the physical systems on the large time scale.
Conservation of the energy, momentum, or the angular momentum in mechanics are the most famous examples of the integrals of motion.
However, the continuous mediums (liquids, gases, solids, plasmas, and the electromagnetic field) have specific integrals of motion as well.
One of them is the helicity $h$, existing in addition to the full energy.
For the classical fluid or gas it is related to the velocity vector field $\textbf{v}$ as the volume integral
$h_{c}=\int (\textbf{v}\cdot [\nabla\times\textbf{v}])d^{3}r$,
where $(\textbf{a}\cdot \textbf{b})$ is the scalar product of two vectors,
and $[\textbf{a}\times \textbf{b}]$ is the vector product of two vectors,
while $\mbox{\boldmath $\Omega$}_{c}=[\nabla\times\textbf{v}]$ is the vorticity of the classical electrically neutral fluid.
The electromagnetic field helicity composed of the vector potential $\textbf{A}$ and the magnetic induction $\textbf{B}$:
$h_{em}=(q/mc)^{2}\int (\textbf{A}\cdot \textbf{B})d^{3}r$
with $\textbf{B}=[\nabla\times\textbf{A}]$
and the electromagnetic vorticity
$\mbox{\boldmath $\Omega$}_{em}=(q/mc)\textbf{B}$,
where
$q$ is the charge of the particle,
$m$ is the mass of  the particle,
and $c$ is the speed of light.
Complete field-material vorticity
$\mbox{\boldmath $\Omega$}_{c}+\mbox{\boldmath $\Omega$}_{em}$
can be found directly from the momentum balance equation
(usually called the Euler equation) of the charged fluid.
Spin density leads to additional material vorticity,
which has quantum nature,
but it is possible to consider it in the quasi-classic model like the Landau--Lifshitz--Gilbert equation.
The spin or quantum vorticity appears independently from the classic and electromagnetic vorticities.
Therefore, its form can be suggested in different forms like $[\nabla\times\textbf{S}]$,
where $\textbf{S}$ is the spin density,
but we expect to consider consistent equations of motions for two types of vorticity to obtain the conserving helicity.
Hence, we follow work \cite{Mahajan PRL 11} and references therein
and present the quantum vorticity
\begin{equation}\label{TCinMF vorticity spin def for Int}
\Omega_{q}^{\alpha}(\textbf{r},t)=\frac{\hbar}{2m}\varepsilon^{\sigma\mu\nu}\varepsilon^{\alpha\beta\gamma}
n^{\sigma}\cdot \partial_{\beta} n^{\mu} \cdot\partial_{\gamma} n^{\nu}, \end{equation}
where
$\hbar$ is the Planck constant,
$\textbf{n}$ is the unit vector field corresponding to the normalized magnetization or the spin density
we also use the tensor notations:
$\alpha$ and other Greek subindexes are the tensor indexes equal to $x,y,z$,
$\varepsilon^{\alpha\beta\gamma}$ is the unit antisymmetric third rank tensor
(Levi-Civita symbol),
and the summation on the repeating indexes is assumed.
For the antiferromagnetic material (AFM), the partial spin vorticity appears for each subspecies,
the classical vorticity appears for each subspecies as well.
Moreover, if the AFM contains some nonmagnetic ions or atoms we need to include the partial classic vorticity.
Combination of all vorticities and their vector potentials enters the full helicity of the system.

Another spin related integral of motion is the topological charge (TC).
This is an integral characteristics of some magnetically nonuniform structures such as the skyrmions.
Substitution of the
anzatz
for the normalized magnetization in the TC definition shows that it is a constant with
integer
value.
Possibility of the
anzatz
application can be related to the dynamical properties of the system
or its static properties related to the balance of forces or the spin torques in the system.
However,
its conservation and the integer value is mostly hidden in the structure of the integral form.
The TC has strong similarity to the spin vorticity.
The "conservation" of the vorticity and the conservation of the hydrodynamic helicity
(its value is usually not discrete or it is not proportional to integer value of a parameter)
is based on the structure of both the forces or the spin torques existing in the system.
It is important to trace the coincidence or possible difference of conditions of their conservation.
Let us also mention the Berry phase \cite{Resta JP CM 00}, \cite{Xiao RMP 10},
which is not considered in this paper,
while it is an important characteristics of some modern materials.

Conservation of the TC is related to its structure in terms of the dimensionless spin density.
However, its integer value is also based on the specific structure of the spin density as well.
This conclusion shows some unification of this notion disconnected with the dynamical properties of the physical system.
So, it leaves open the formal question of the possibility of creation of the structures with the topological charge (TC).
They are expected to be created at the conservation of the full TC in the sample,
but it leads to the formation opposite topological charges in close proximity to each other.
So, some distribution of the TC is formed together with the TC density (TCD).

The interaction under consideration in this paper, usually modeling the behavior of multiferroic materials,
show the conservation of the TC and the hydrodynamic helicity being the integral characteristics of the physical system,
but the spin vorticity, being a local characteristics of the magnetically ordered material,
shows some violation, which is compensated with the phonon dynamics.
It shows some essential interplay between magnons and phonons even if we do not consider the magnetoelectric coupling,
where the electric part is usually related to the deformation.
The damping violate the energy conservation and affects the helicity evolution as well,
but we do not consider it in this paper.

Necessary condition for the solid sample stability is the zero value of the velocity field in the equilibrium regime.
This concept of the polarization formation is considered in Ref. \cite{Hu PRL 08}.
Later, it is reviewed in Ref. \cite{AndreevTrukh JETP 24}.
This concept is physically reasonable,
but it does not allow to obtain the well-known spin-current model
$P^{\mu}
\sim\varepsilon^{\mu\alpha\beta}J^{\alpha\beta}$
suggested in Ref. \cite{Katsura PRL 05}.
Since the spin current is not involved in the model considered in Ref. \cite{Hu PRL 08}.
However, to some extend, this model \cite{Hu PRL 08} reproduce the relationship between the polarization and magnetization
considered by Mostovoy in Ref. \cite{Mostovoy PRL 06}.
While the polarization from Ref. \cite{Mostovoy PRL 06} can be found from the spin-current model
\cite{Katsura PRL 05} and \cite{Tokura RPP 14}.
A phenomenological analysis of the polarization based mostly on the symmetry is presented in review \cite{Dong AinP 15}.

The spin-current model is suggested for the particular mechanism of the polarization formation \cite{Katsura PRL 05}.
However, in Ref. \cite{AndreevTrukh JETP 24},
it is demonstrated that the electric polarization can be caused with the spin currents of different nature.
Consequently, different types of the spin currents give different form of the electric polarization \cite{AndreevTrukh JETP 24}.
Particularly, in Ref. \cite{AndreevTrukh JETP 24}, three mechanisms of the polarization formation are discussed.
One gives the polarization considered by Mostovoy in Ref. \cite{Mostovoy PRL 06} associated with noncollinear spins.
The second mechanism gives the polarization associated with collinear parts of spins
(being in the system of partially noncollinear spins).
The third mechanism has a kinetically-quantum nature and the related to the contribution of the quantum Bohm potential in the spin evolution equation.
The resulting polarization is similar to the polarization considered by Mostovoy in Ref. \cite{Mostovoy PRL 06}
if density is uniform,
but this mechanism gives additional contribution for the nonuniform systems or in the dynamical regimes of the longitudinal phonon propagation.
Described generalized spin-current model is originally developed for the ferromagnetic spin order \cite{AndreevTrukh JETP 24},
but later it is generalized on the AFM spin order.
In this paper, we point out that
the spin-current model existence is in consistency
with the conservation of main characteristics analyzed in this paper:
the TC, the vorticity, and the helicity.
Moreover, the conservation of these characteristics require appearance of additional models
for the appearance of the deformation (and corresponding polarization) of spin origin,
and in reverse, the appearance of the magnetization
(nonuniform magnetization via creation of the corresponding effective spin current)
of electric polarization origin.
They are called as
the reverse spin-current model of magnetization origin from the electric polarization, and
the spin-field model of the electric polarization of the spin origin.

In context of the described program we discuss in the text major consequencies of the spin-current model
and its relations with the symmetric Heisenberg exchange interaction and
the Dzylaoshinskii-Moriya interaction.
Moreover, the novel form of the spin-spin interaction related to the odd anisotropy of the symmetric
exchange interaction (OASEI) is discussed following Ref. \cite{Andreev 2025 12}.
Appearance of concept of OASEI is motivated by the possibility of the
Keffer-like form of the symmetric Heisenberg exchange integral.
The OASEI exists in AFM similarly to the
Dzylaoshinskii-Moriya interaction of weak ferromagnetic type for AFM.
We also discuss the additional spin caused polarization formed by
the OASEI like it is suggested in Ref. \cite{Andreev 2025 12}.

This paper is organized as follows.
In Sec. II the notion of the topological charge is discussed and its relation to the spin vorticity is shown.
In Sec. III the spin vorticity evolution is considered, its "nonconservation" is demonstrated, and the souses of the spin vorticity is obtained.
In Sec. IV conservation of the topological charge
and cancelation of the spin vorticity souses is described.
In Sec. V the full vorticity conservation and
the hydrodynamic helicity
conservation including the evolution of the classic vorticity following from the Euler equation is discussed.
In Sec. VI the spin-current model is discussed as the background to development of the reverse spin-current model.
In Sec. VII the reverse spin-current model is suggested to demonstrate the magnetization and
the spin-current origin from the electric polarization.
In Sec. VIII the spin-field model of the electric polarization is suggested in addition to the spin-current model.
In Sec. IX a brief summary of obtained results is presented.

\section{Topological charge and spin vorticity}

Topological charge can be considered as the surface integral of the complex structure of the normalized spin density $\textbf{n}$
\begin{equation}\label{TCinMF Topological charge xy}
q_{T}=\frac{1}{4\pi}
\int (\textbf{n}\cdot [\partial_{x} \textbf{n} \times\partial_{y} \textbf{n}]) dxdy, \end{equation}
for the plane magnetic films
(see \cite{Rybakov PRB 19}).
Here the normalized spin density can be reproduced via the full local magnetization $\textbf{M}$,
the magnetic moment of the ion/atom $\mu$ and concentration $n$:
$\textbf{n}=\textbf{M}/\mu n$.

It is useful to consider the integral over arbitrary surface, for the more general analysis,
$$q_{T}=\frac{1}{4\pi}\varepsilon^{\alpha\beta\gamma}
\int(\textbf{n}\cdot [\partial_{\beta} \textbf{n} \times\partial_{\gamma} \textbf{n}]) dS^{\alpha}$$
\begin{equation}\label{TCinMF Topological charge arb surface}
=\frac{1}{4\pi}\varepsilon^{\alpha\beta\gamma}
\int n^{\alpha} ([\nabla n^{\beta} \times\nabla n^{\gamma}] \cdot d\textbf{S}), \end{equation}
where $d\textbf{S}$ or $dS^{\alpha}$ is the element of the surface including the normal direction to the surface.
The first of the presented forms is made in the vector form for the normalized spin density
and its derivatives composed into the triple scalar product,
while the derivatives and the element of surface also composed into the triple scalar product,
but it is written in the tensor notations.
The second of presented forms
is made
vice versa.

Basically, equation (\ref{TCinMF Topological charge xy}) presents the surface integral of the z-projection of the spin vorticity:
\begin{equation}\label{TCinMF vorticity spin def}
\Omega_{q}^{\alpha}(\textbf{r},t)=\frac{\hbar}{2m}\varepsilon^{\sigma\mu\nu}\varepsilon^{\alpha\beta\gamma}
n^{\sigma}\cdot \partial_{\beta} n^{\mu} \cdot\partial_{\gamma} n^{\nu}. \end{equation}



Comparison of equations
(\ref{TCinMF Topological charge arb surface})
and
(\ref{TCinMF vorticity spin def})
shows close relation between
the topological charge (or its density)
(\ref{TCinMF Topological charge arb surface})
and
the hydrodynamic spin vorticity
(\ref{TCinMF vorticity spin def})
$$q_{T}=\int \varrho_{T}dV=\frac{1}{4\pi}\frac{2m}{\hbar}
\int (\mbox{\boldmath $\Omega$}_{q}\cdot d\textbf{S})$$
\begin{equation}\label{TCinMF Topological charge via vort}
=\frac{1}{4\pi}\frac{2m}{\hbar}
\int (\nabla\cdot\mbox{\boldmath $\Omega$}_{q}) dV, \end{equation}
where
the divergence theorem is used.
Hence, we can present the form for the topological charge density
\begin{equation}\label{TCinMF TC density via vort}
\varrho_{T}=
\frac{1}{4\pi}\frac{2m}{\hbar}
(\nabla\cdot\mbox{\boldmath $\Omega$}_{q}).
\end{equation}
If the sample has the full local spin polarization $\textbf{n}^2=1$ we get $(\nabla\cdot\mbox{\boldmath $\Omega$}_{q})=0$.
Same is true if we have system with partial spin polarization, but the value of spin polarization is fixed over the sample
$\textbf{n}^2=\eta^{2}=const\leq 1$.


Another form of
the topological charge density
can be found in literature
\cite{Yang PRA 08}, \cite{Girvin PT 00}, \cite{Moon PRB 95}):
\begin{equation}\label{TCinMF Topological charge density lit}\varrho_{T}(\textbf{r})=\frac{1}{8\pi}
\varepsilon^{ij}\textbf{s}\cdot \partial_{i} \textbf{s} \times\partial_{j} \textbf{s}. \end{equation}
The topological charge is an invariant for two component Bose-Einstein condensates \cite{Yang PRA 08}.
Some hydrodynamic properties of spin-1 Bose-Einstein condensates are discussed in Ref. \cite{Andreev 2024 Ph B}
with the stress on the vortical properties.


\subsection{On the integer value of the topological charge}

The spin or magnetization field can be considered as the unit field for the ferromagnetic phase
(neglecting the decrease of the magnetization with the growth of temperature close to the Curie temperature)
$$\textbf{n}(\textbf{r},t)$$
\begin{equation}\label{BECTSpin1Skyrmion s unit vector}
=\{\cos\Phi(\textbf{r})\cdot \sin\Theta(\textbf{r}), \sin\Phi(\textbf{r})\cdot \sin\Theta(\textbf{r}), \cos\Theta(\textbf{r})\}. \end{equation}
Functions $\Phi(\textbf{r},t)$ and $\Theta(\textbf{r},t)$ depend in general on the three dimensional coordinate $\textbf{r}$.

We can focus on the plane samples.
We choose the $z$ axis perpendicular to the plane of symmetry and
we get no dependence of functions $\Phi$ and $\Theta$ on coordinate $z$.
We choose the polar coordinates in $x-y$ plane.
Hence, we use $x=\varrho\cos\varphi$ and $y=\varrho\sin\varphi$.
Moreover, function $\Phi(\textbf{r})$ can be reduced to $\pm n\varphi$ at the analysis of the skyrmion solutions
\cite{Bera PRR 19}, \cite{Leonov NJP 16}.
To obtain, for instance, the skyrmion solution function $\Theta(\textbf{r})$ can be considered as the function
of radial polar coordinate $\Theta(\textbf{r})=\Theta(\varrho)$,
where $\varrho=\sqrt{x^2+y^2}$.
Hence, equation (\ref{BECTSpin1Skyrmion s unit vector}) is reduced to
\begin{equation}\label{BECTSpin1Skyrmion s unit vector reduced}
\textbf{n}=\{\cos n\varphi \sin\Theta(\varrho), \sin n\varphi \sin\Theta(\varrho), \cos\Theta(\varrho)\}. \end{equation}
This assumption leads to the integer value of the TC $q_{T}=n$.

\section{The spin vorticity evolution under different interactions in magnetically ordered materials}

Let us consider the spin vorticity evolution equation under action of a number of interactions usually considered for the magnetically ordered materials.
Mainly we are focused on the anisotropy energy, exchange (symmetric) interaction,
magnetic dipole-dipole interaction, DMI, spin-orbit interaction.
Majority of them give sources of the spin vorticity.
But, it is necessary to include the classical vorticity related to the velocity field and deformation of the medium
(the contribution of phonons in the dynamics of the system).
So, the complete vorticity has no spin-torque or spin related force souses
(the spin-orbit interaction gives few terms for the vorticity sources,
but they can be excluded using the spin-current model of the electric polarization in the multiferroics).
It also leads to the conservation of the hydrodynamic helicity.

\subsection{Dipole-dipole interaction of the magnetic moments and the quantum Bohm potential}

We start analysis of the structure of the spin vorticity evolution equation
with the consideration of the dipole-dipole interaction between the magnetic moments.
Hence, the partial spin evolution equation can be presented as
\begin{equation}\label{TCinMF spin evol with DDI}
\partial_{t}\textbf{n}=
\gamma[\textbf{n}\times\widehat{\textbf{B}}], \end{equation}
where
the effective magnetic field
\begin{equation}\label{TCinMF magn field eff in vort}
\widehat{\textbf{B}}=\textbf{B}
+\frac{\hbar }{2m \gamma}\frac{1}{n}\partial^{\beta}(n\partial^{\beta}\textbf{n})\end{equation}
includes the quantum spin current (the contribution of the quantum Bohm potential)
along with the induction of the magnetic field $\textbf{B}$.
While the magnetic field induction $\textbf{B}$ satisfies the following field equations
\begin{equation}\label{TCinMF }
(\nabla\cdot\textbf{B})=0,
 \end{equation}
and
\begin{equation}\label{TCinMF }
[\nabla\times\textbf{B}]=4\pi[\nabla\times(\mu n\textbf{n})], \end{equation}
where $\mu$ is the magnetic moment of the atom/ion,
and $\textbf{M}=\mu n\textbf{n}$ is the full magnetization,
here we approximately consider that
the magnetization module $\mid\textbf{M}\mid=M_{s}$ is equal to the saturation magnetization $M_{s}$.

Spin vorticity (\ref{TCinMF vorticity spin def}) satisfies the following evolution equation
\begin{equation}\label{TCinMF vort evol Q with DDI}
\partial_{t}\mbox{\boldmath $\Omega$}_{q}=
\nabla\times(\textbf{v}\times\mbox{\boldmath $\Omega$}_{q})
+\gamma\nabla n^{\beta}\times\nabla \widehat{B}^{\beta}, \end{equation}
if the spin evolve in accordance with equation (\ref{TCinMF spin evol with DDI}) \cite{Mahajan PRL 11}
(actually equation (\ref{TCinMF vort evol Q with DDI}) includes the convective part of the spin evolution,
which is not presented explicitly in equation (\ref{TCinMF spin evol with DDI})).
First two terms in equation (\ref{TCinMF vort evol Q with DDI}) represent the well-known structure for the vorticity evolution equation,
while the last
term $\gamma\nabla n^{\beta}\times\nabla \widehat{B}^{\beta}$
appears to be the source of the spin vorticity:
$\mbox{\boldmath $\Sigma$}_{\Omega}=\gamma\nabla n^{\beta}\times\nabla \widehat{B}^{\beta}$.
In this paper we deal with the ferromagnetic objects.
Therefore, we need to consider behavior of other spin torques entering the spin evolution equation
and their contribution in the source of the vorticity.
It is well known that the majority of the spin torques in
the Landau--Lifshitz--Gilbert equation
can be considered as the effective magnetic field.
Hence, addition terms conserve structure of the spin evolution equation (\ref{TCinMF spin evol with DDI})
and the vorticity evolution equation (\ref{TCinMF vort evol Q with DDI}).
Nevertheless, we discuss main contributions in the Landau--Lifshitz--Gilbert equation with known microscopic interactions.
Presence of the vorticity sources leads to the hydrodynamic helicity nonconservation.
Classical vorticity $\mbox{\boldmath $\Omega$}_{q}=\nabla\times\textbf{v}$ can be combined with the quantum vorticity is the full vorticity,
so it leads to the conservation of the full hydrodynamic helicity \cite{Mahajan PRL 11}.


Below we suggest the reverse spin-current model,
which provides the dynamical mechanism based on the spin-orbit interaction
leading to the contribution of the electric polarization in the spin evolution equation.
However, the kinematic contribution of the polarization in the spin evolution equation
and the vorticity evolution equation is possible as well.
Hence we mention it here.

Both equations (\ref{TCinMF spin evol with DDI}) and (\ref{TCinMF magn field eff in vort})
neglect the convective part of the time derivative of the normalized spin $\textbf{n}$
(it is proportional to $(\textbf{v}\cdot\nabla)\textbf{n}$).
If we remember that
the velocity field is proportional to the time evolution of the deformation $\textbf{v}=\dot{\textbf{u}}$.
So, it does not require some flows in systems like in the fluids or the gases.
In the same way we can consider the momentum evolution in the Euler equation
$(\textbf{v}\cdot\nabla)\textbf{v}= (\dot{\textbf{u}}\cdot\nabla)\dot{\textbf{u}}$.
Moreover, the deformation $\textbf{u}$ is related to the electric polarization in the system of ions.
Therefore, the convective part of the time derivative is an example of
the magnetization-electric polarization "interaction"
(or some kinematic interplay of these factors).
We also neglect the thermal part of the spin-current or degeneracy part of the spin-current (related to the Pauli blocking).
First is neglegible for the small temperatures,
while the second is neglegible for the relatively heavy ions
(being essential for the electrons in the conducting mediums).

All described above shows that the temporal dynamics of deformations and electric polarization can be considered
as the part of
the vorticity evolution equation
in the absence of any flux
\begin{equation}\label{TCinMF vort evol Q with DDI 2 with deform}
\partial_{t}\mbox{\boldmath $\Omega$}_{q}=
\nabla\times(\dot{\textbf{u}}\times\mbox{\boldmath $\Omega$}_{q})
+\gamma\nabla n^{\beta}\times\nabla \widehat{B}^{\beta}. \end{equation}

\subsection{Anisotropy energy associated torque}

Most interactions in the spin torque for the magnetically ordered materials can be considered as the result of action of the effective magnetic field.
Therefore, equation for the spin vorticity evolving under the DDI (\ref{TCinMF vort evol Q with DDI}) is highly useful for the further application to other interactions.

In this case we get the effective magnetic field $\textbf{B}_{eff,an}=\kappa S_{z}\textbf{e}_{z}/\gamma$
So, the last term in equation (\ref{TCinMF vort evol Q with DDI}) equals to zero
$\mbox{\boldmath $\Sigma$}_{\Omega}=\gamma\nabla n^{\beta}\times\nabla B^{\beta}_{eff}=\kappa \nabla n^{z}\times\nabla S_{z}=0$
for the uniform concentration $n=const$ with the spin density $S_{z}=m_{s}\hbar n n_{z}$
and the maximal spin projection $m_{s}\hbar$.

So, this contribution approximately equals to zero,
in the approximation of constant density.
However, this vorticity source can be compensated from the contribution of the classical vorticity.

\subsection{Symmetric Heisenberg exchange interaction}

Here we consider the contribution of the symmetric Heisenberg exchange interaction
$\hat{H}=U(r_{12})(\hat{\textbf{S}}_{1}\cdot \hat{\textbf{S}}_{2})$
in the effective magnetic field $\textbf{B}_{eff,ex}=A \triangle\textbf{S}/\gamma$,
with $A=g_{u}/6$
represented via the interaction constant
$g_{u}=\int r^2 U(r)d^{3}r$
being the integral of the exchange integral.
It leads to nonzero source of the quantum vorticity related to the symmetric Heisenberg exchange interaction
\begin{equation}\label{TCinMF }
\mbox{\boldmath $\Sigma$}_{\Omega}=A\nabla n^{\beta}\times\nabla \triangle S^{\beta}\neq0
. \end{equation}

Further analysis of the Euler equation for the velocity field associated with the classical vorticity
shows that
$\textbf{B}_{eff,ex}=A \triangle\textbf{S}/\gamma$ is a part of the effective magnetic field
associated with the symmetric Heisenberg exchange interaction.
In contrast with the anisotropy energy we can get incomplete cancelation of the vorticity sources
(or their approximate cancelation).

\subsection{Dzylaoshinskii-Moriya interaction of helicoid type}

There is a number of the Dzylaoshinskii-Moriya interaction torques being related to the same microscopic Hamiltonian of form of
$\textbf{D}_{12}\cdot [\hat{\textbf{S}}_{1}\times \hat{\textbf{S}}_{2}]$.
Their macroscopic form is related to the structure of the Dzylaoshinskii vector constant $\textbf{D}_{12}$
(see corresponding discussion in Ref. \cite{Andreev 2025 11}).
In this paper we start discussion of the Dzylaoshinskii-Moriya interaction torques with the regime we called the helicoid type
since it can favor the inhomogeneos spin structures in form of helix.

These structures can be described
analytically as the noncollinear periodically changing spin density structures
\begin{equation}\label{TCinMF helix}
\textbf{S}_{0}=S_{b}\cos(qz)\textbf{e}_{x}+S_{c}\sin(qz)\textbf{e}_{y},
\end{equation}
where the trigonometric functions depend on the direction $z$ perpendicular to the nonzero spin projections.
Here $S_{b}$ and $S_{c}$ are the partial amplitudes of the spin density,
$\textbf{e}_{x}$ and $\textbf{e}_{y}$ are the unit vectors in Cartesian coordinates,
$z$ is the coordinates in the corresponding direction,
$q$ is the wave vector of the equilibrium spin structure.

Corresponding Dzylaoshinskii-Moriya interaction torques is
\begin{equation}\label{TCinMF s evolution for n DMI helix}
\partial_{t}\textbf{n}_{DMIh}=
\frac{1}{3}g_{(\gamma)}[\textbf{n}\times curl\textbf{S}],
\end{equation}
where we have the following interaction constant related to function
$\gamma(r_{ij})$ in the Dzyaloshinskii constant $\textbf{D}_{ij}=\gamma(r_{ij})\textbf{r}_{ij}$:
$g_{(\gamma)}=\int r^2 \gamma(r)d^{3}r$.

The Dzylaoshinskii-Moriya interaction of helicoid type leads to the following
effective magnetic field $\textbf{B}_{eff,DMIh}=\frac{1}{3}g_{(\gamma)} curl\textbf{S}/\gamma$
providing the source of the vorticity
\begin{equation}\label{TCinMF }
\mbox{\boldmath $\Sigma$}_{\Omega}=\frac{1}{3}g_{(\gamma)} \nabla n^{\beta}\times\nabla (\varepsilon^{\beta\mu\nu}\partial^{\mu}S^{\nu})
. \end{equation}

\subsection{Dzylaoshinskii-Moriya interaction of cycloid type}

There is the noncollinear periodical spin structures
\begin{equation}\label{TCinMF spin cycloid}
\textbf{S}_{0}=S_{b}\cos(qx)\textbf{e}_{x}+S_{c}\sin(qx)\textbf{e}_{y}
\end{equation}
called the spin cycloid,
where the direction of the chain is in the plane of spin projections.
The formation of this type of structures is related to the Dzylaoshinskii-Moriya interaction.
However, this form the Dzylaoshinskii-Moriya interaction is related to the Keffer form of the Dzylaoshinskii constant \cite{Khomskii JETP 21},
which includes the vector of shift of the nonmagnetic ion (ligand) placed between two interacting magnetic ions.
In equation (\ref{TCinMF spin cycloid}) we have the following notations:
$S_{b}$ and $S_{c}$ are the partial amplitudes of the spin density,
$\textbf{e}_{x}$ and $\textbf{e}_{y}$ are the unit vectors in Cartesian coordinates,
$x$ is the coordinates in the corresponding direction,
$q$ is the wave vector of the equilibrium spin structure.

In described regime the spin torque has the following for at the macroscopic description
$$\partial_{t}\textbf{n}_{DMIc}=-\frac{1}{3}g_{(\beta)}[\textbf{n}\times[\tilde{\nabla}\times\textbf{S}]]$$
\begin{equation}\label{TCinMF s evolution for n DMI cycloid}
=\frac{1}{3}g_{(\beta)}\biggl((\textbf{n}\cdot[\mbox{\boldmath $\delta$}\times\nabla])\textbf{S}
-n^{\beta}[\mbox{\boldmath $\delta$}\times\nabla]S^{\beta}\biggr),
\end{equation}
where
$\tilde{\nabla}\equiv[\mbox{\boldmath $\delta$}\times\nabla]$,
so we get the effective $curl_{eff}\textbf{S} \equiv[\tilde{\nabla}\times\textbf{S}]$.
So, equation (\ref{TCinMF s evolution for n DMI cycloid})
can be rewritten in the form similar to equation (\ref{TCinMF s evolution for n DMI helix}):
$\partial_{t}\textbf{n}_{DMIc}=-\frac{1}{3}g_{(\beta)}[\textbf{n}\times curl_{eff}\textbf{S}$.
We also have the following interaction constant related to function
$\beta(r_{ij})$ in the Dzyaloshinskii constant $\textbf{D}_{ij}=\beta(r_{ij})[\textbf{r}_{ij}\times\mbox{\boldmath $\delta$}]$:
$g_{(\beta)}=\int r^2 \beta(r)d^{3}r$.

The spin evolution equation (\ref{TCinMF s evolution for n DMI cycloid}) allows to introduce
the effective magnetic field $\textbf{B}_{eff,DMIc}= -  \frac{1}{3}g_{(\beta)} [\tilde{\nabla}\times\textbf{S}]/\gamma$,
which leads to the spin vorticity source
\begin{equation}\label{TCinMF }
\mbox{\boldmath $\Sigma$}_{\Omega}= - \frac{1}{3}g_{(\beta)} \nabla n^{\beta}\times\nabla (\varepsilon^{\beta\mu\nu}\tilde{\nabla}^{\mu}S^{\nu})
. \end{equation}


\subsection{Dzylaoshinskii-Moriya interaction of weak ferromagnetic type for AFM}

\begin{figure}\includegraphics[width=8cm,angle=0]{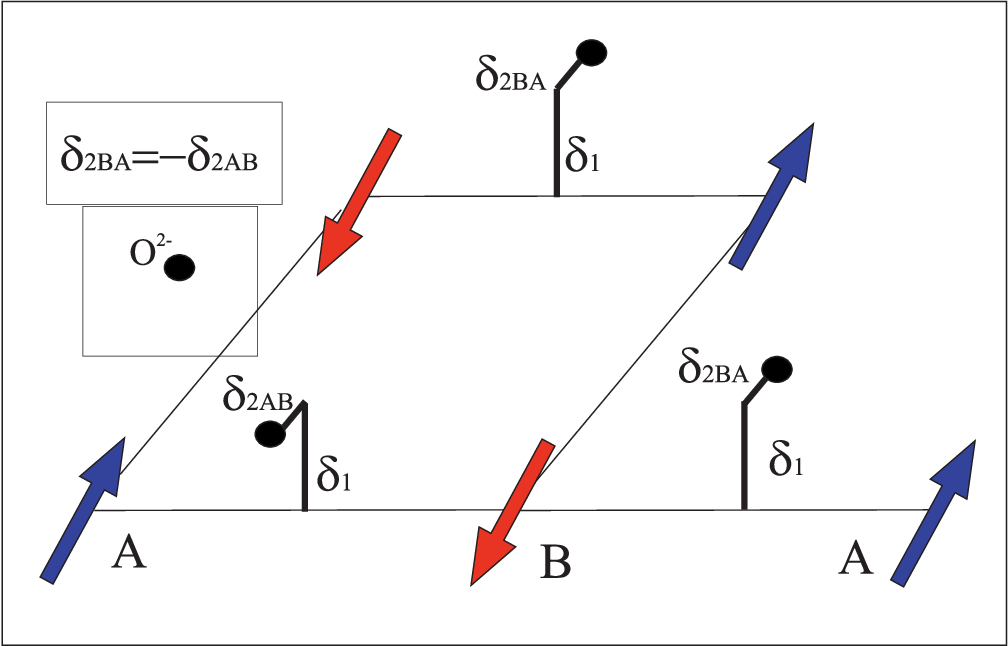}
\caption{\label{TCinMF Fig 01}
Partial ligand shifts are illustrated.
The magnetic ions are presented within the red and blue arrows
(spin-up type A and spin-down configurations type B).
It illustrates necessary elements for
the Dzylaoshinskii-Moriya interaction of weak ferromagnetic type for AFM
and
the symmetric
exchange interaction with the odd anisotropy.}\end{figure}

Two Dzylaoshinskii-Moriya interactions described above can be considered for different magnetically ordered structures
including the ferromagnetic materials and periodic structures in ferromagnetic materials.
Here, we focus on the Dzylaoshinskii-Moriya interactions existing in two component magnetic structures such as antiferromagnetic materials.
This type of Dzylaoshinskii-Moriya interaction is responsible for the weak ferromagnetism in the antiferromagnetic materials.
Its appearance is related to the reverse magnetoelectric effect,
since its existence is related to the electric polarization (the ligand shift).
To some extend there is similarity to the regime (\ref{TCinMF s evolution for n DMI cycloid})  based on the Keffer form for the Dzylaoshinskii constant,
but here we need more like quandrupolar structure of the electric dipoles presented in Fig. (\ref{TCinMF Fig 01}).

It leads to the spin torques acting on subspecies $A$ ("spin-up")
\begin{equation}\label{TCinMF s evolution DMI wf A}
\partial_{t}\textbf{n}_{A,DMI}=
-g_{(0\zeta_{1})}[\textbf{n}_{A}\times [\mbox{\boldmath $\delta$}_{2,AB} \times \textbf{S}_{B}]],
\end{equation}
and $B$ ("spin-down")
\begin{equation}\label{TCinMF s evolution DMI wf B}
\partial_{t}\textbf{n}_{B,DMI}=
g_{(0\zeta_{1})}[\textbf{n}_{B}\times [\mbox{\boldmath $\delta$}_{2,AB} \times \textbf{S}_{A}]],
\end{equation}
where
$g_{(0\zeta_{1})}$
is the corresponding interaction constant,
it is similar to the interaction constant presented above,
it is related to the coefficient being function of the relative coordinate of the magnetic ions in the microscopic Hamiltonian,
it sometimes is presented in literature as $D$ similarly to the interaction constants in other forms of the Dzylaoshinskii-Moriya interaction considered above.
Equations (\ref{TCinMF s evolution DMI wf A}) and (\ref{TCinMF s evolution DMI wf B})
contains the partial shift of the ligand $\mbox{\boldmath $\delta$}_{2,AB}$.

Equations (\ref{TCinMF s evolution DMI wf A}) and (\ref{TCinMF s evolution DMI wf B})
allow to introduce
the effective magnetic field acting on subspecies $A$:
$\textbf{B}_{eff,DMI,A}=-g_{(0\zeta_{1})} [\mbox{\boldmath $\delta$}_{2,AB} \times \textbf{S}_{B}]/\gamma$,
which leads to the source of the spin vorticity
\begin{equation}\label{TCinMF }
\mbox{\boldmath $\Sigma$}_{\Omega}=-g_{(0\zeta_{1})} \nabla n^{\beta}\times\nabla (\varepsilon^{\beta\mu\nu}\delta^{\mu}_{2,AB}S_{B}^{\nu})
. \end{equation}

\subsection{The symmetric
exchange interaction with the odd anisotropy}

The Keffer-like form of the symmetric Heisenberg exchange integral can be considered in AFM \cite{Andreev 2025 12}.
It gives the novel form of the spin-spin interaction caused by
the odd anisotropy
the symmetric
exchange interaction (OASEI),
where the exchange integral has the following form
\begin{equation}\label{TCinMF U structure in Ham HHI 2}
U_{2,ij}=l_{ij}(\textbf{r}_{ij}\cdot[\mbox{\boldmath $\delta$}_{1}\times \mbox{\boldmath $\delta$}_{2,ij-AB}]).
\end{equation}

The OASEI gives the following contribution in the Landau--Lifshitz--Gilbert equation for subspecies $A$
\begin{equation}\label{TCinMF S A evol OASEI}
\partial_{t}\textbf{n}_{A}=-\frac{1}{3}g_{(l)}[\textbf{n}_{A}\times(\mbox{\boldmath $\delta$}_{eff}\cdot\nabla)\textbf{S}_{B}]
\end{equation}
and, similarly for subspecies $B$
\begin{equation}\label{TCinMF S B evol OASEI}
\partial_{t}\textbf{n}_{B}=\frac{1}{3}g_{(l)}[\textbf{n}_{B}\times(\mbox{\boldmath $\delta$}_{eff}\cdot\nabla)\textbf{S}_{A}].
\end{equation}
where $\mbox{\boldmath $\delta$}_{eff}=\mbox{\boldmath $\delta$}_{1}\times \mbox{\boldmath $\delta$}_{2,AB}$
is the combination of the partial shifts of the ligand (see Fig. \ref{TCinMF Fig 01}).

In this regime we can present
the effective magnetic field acting on subspecies $A$
\begin{equation}\label{TCinMF }
\textbf{B}_{eff,OASEI,A}=-\frac{1}{3}g_{(l)} (\mbox{\boldmath $\delta$}_{eff}\cdot\nabla)\textbf{S}_{B}/\gamma
. \end{equation}
It gives the source of the spin vorticity
\begin{equation}\label{TCinMF }
\mbox{\boldmath $\Sigma$}_{\Omega}=-\frac{1}{3}g_{(l)} \nabla n^{\beta}\times\nabla ((\mbox{\boldmath $\delta$}_{eff}\cdot\nabla)S_{B}^{\beta})
. \end{equation}

Below we consider the spin-current model of the electric polarization,
which can include the interaction considered in this subsection.
It leads to the polarization of form of
\begin{equation}\label{TCinMF Pol nm 1}
\textbf{P}=\frac{1}{3}\frac{\gamma}{c}g_{(l)}
[\textbf{S}_{A}(\textbf{S}_{B}\cdot\mbox{\boldmath $\delta$}_{eff})
-\textbf{S}_{B}(\textbf{S}_{A}\cdot\mbox{\boldmath $\delta$}_{eff})].
\end{equation}

\subsection{Spin-orbit interaction}

The spin-orbit interaction plays major role for both the spin-current model of the electric polarization in muliferroics derivation
and the reverse spin-current model of the magnetization of the electric polarized mediums.
Here we consider the spin-orbit interaction related spin torque
\begin{equation}\label{TCinMF s evolution SOI long H}
\partial_{t}\textbf{n}=\frac{1}{n}\textbf{T}_{SO}, \end{equation}
where the spin torque can be presented in the tensor form only
\begin{equation}\label{TCinMF}T^{\alpha}_{SO}=- \frac{2\mu}{\hbar c}
\varepsilon^{\alpha\beta\gamma}\varepsilon^{\beta\mu\nu}E^{\mu} J^{\gamma\nu}
,\end{equation}
since it includes the spin current tensor.
We do not consider here the part of the kinetic spin current related to the spin-orbit interaction.


If we consider the kinetic spin current $J^{\gamma\nu}=S^{\gamma}v^{\nu}$
we can introduce the effective magnetic field as well.
In this case the spin torque simplifies to
$T^{\alpha}_{SO}=\frac{2\mu}{\hbar c}\varepsilon^{\alpha\beta\gamma}\varepsilon^{\gamma\mu\nu}E^{\mu} S^{\beta}v^{\nu}$
and we obtain
$\textbf{B}_{eff}=\frac{2\mu}{\hbar c}[\textbf{E}\times \textbf{v}]/\gamma$.

Hence it gives the following source of the spin vorticity
\begin{equation}\label{TCinMF }
\mbox{\boldmath $\Sigma$}_{\Omega}=\frac{2\mu}{\hbar c} \nabla n^{\beta}\times\nabla ([\textbf{E}\times \textbf{v}]^{\beta})
. \end{equation}

We consider crystal structures, where no flow of particles is possible,
but the velocity field $\textbf{v}$ is related to the dynamics of the deformation $\dot{\textbf{u}}$ or the evolution of polarization
\begin{equation}\label{TCinMF }
\mbox{\boldmath $\Sigma$}_{\Omega}=\frac{2\mu}{\hbar c q_{eff}n} \nabla n^{\beta}\times\nabla ([\textbf{E}\times \dot{\textbf{P}}]^{\beta})
. \end{equation}
The replacement of the velocity field $\textbf{v}$ within the deformation time derivative $\dot{\textbf{u}}$
has no relation to the source of the deformation,
so this representation can be applied to multiferroics of all types,
including the second type multiferroics,
where the electric polarization is caused by the spin distribution.

Further analysis of its contribution is presented below
after introduction of the reverse spin-current model for the electric polarization-magnetization relation.

\subsection{Magneto-electric effect contribution}

The contribution of the magneto-electric effect in the Landau--Lifshitz--Gilbert equation appears
via the spin-current model discussed below.
However, we consider this effect here for one of several regimes appearing in the spin-current model.
We focus on the spin torque appearing due to the electric polarization of spin origin associated with the collinear part of interacting spins.
It is also discussed in Ref. \cite{Andreev 2025 11} (see eqs. (40)-(42)) including the AFM,
and the electric polarization of spin origin associated with the noncollinear part of interacting spins (see eqs. (34)-(36)) including the AFM.

In the chosen regime we find the following expression for the spin torque in the ferromagnetic multiferoics
\begin{equation}\label{TCinMF s evolution MEE col}
\partial_{t}\textbf{n}=
2 c_{2}[\textbf{n}\times
\{(\mbox{\boldmath $\delta$}\cdot \textbf{E})\triangle \textbf{S}
+(\mbox{\boldmath $\delta$}\cdot(\partial^{\delta} \textbf{E}))\cdot\partial^{\delta} \textbf{S}\}].
\end{equation}
It gives the effective magnetic field
\begin{equation}\label{TCinMF }
\textbf{B}_{eff}=2 c_{2} \partial^{\delta}[(\mbox{\boldmath $\delta$}\cdot \textbf{E})\cdot\partial^{\delta} \textbf{S}]/\gamma
, \end{equation}
which leads to the spin vorticity source
\begin{equation}\label{TCinMF }
\mbox{\boldmath $\Sigma$}_{\Omega}=2 c_{2} \nabla n^{\beta}\times\nabla (\partial^{\delta}[(\mbox{\boldmath $\delta$}\cdot \textbf{E})\cdot\partial^{\delta} S^{\beta}])
. \end{equation}

Described here magnetoelectric contribution in the spin evolution equation appears from the spin-current model,
while the spin-current model appears from the Euler equation for the velocity field
due to the compensation of the electric dipole-dipole interaction term.
So, we assume that similar magneto-electric contribution does not appear in the Euler equation.
So, presented source of the spin vorticity is not compensated at the consideration of the full vorticity and hydrodynamic helicity as well.
However, its structure similar to the effective magnetic field corresponds to the conservation of the TC.


\subsection{The Gilbert damping}

The Gilbert damping of the magnetic excitations is related to the interaction of the magnetic part of the system with the nonmagnetic degrees of freedom
\begin{equation}\label{MFMemf s evolution MAIN TEXT}
\partial_{t}\textbf{n}=
a[\textbf{n}\times\partial_{t}\textbf{S}], \end{equation}
with $a<0$.
Energy loss also leads to the nonconservation of the vorticity and the hydrodynamic helicity
This term can be considered as the contribution of the effective magnetic field as well:
$\textbf{B}_{eff,GD}=a \partial_{t}\textbf{S}/\gamma$.
This term has no immediate counterpart in the Euler equation considered below.
The spin-orbit interaction gives the force field containing one term proportional to the time derivative,
so they can contribute in the possibility of cancelation of this vorticity source,
but the spin-orbit interaction contains the contribution of the electric field.
In some cases this field can be replace effective average field related to the distribution of ions.

\section{Topological charge density evolution equation}

Consideration of the TC inside the closed surface formally allows to introduce the TC density as the divergence of the spin vorticity,
which is equal to zero.
To some extend, it shows a reasonable physical property:
parts of the spin structure has no TC or its fraction,
while whole object has a TC.
On the other hand, it puts a restriction on the force field structure (and corresponding source of the vorticity).
Here, we can consider divergence of the spin vorticity evolution equation and consider the behavior of the vorticity sources.

Majority of the considered interactions create sources of the spin vorticity.
We can use equations presented above for the analysis of the equation for the topological charge density evolution.


\section{Full vorticity "conservation" and the hydrodynamic helicity conservation}

Conservation of the topological charge under considered interactions is the essential feature of the topological charge as an integral of motion of the magnetically ordered medium.
However, there is another integral of motion for the spatially distributed systems:
the hydrodynamic helicity.
It is closely related to the vorticity,
which can be represented via the vector potential
$\mbox{\boldmath $\Omega$}=\nabla\times\mathcal{A}$
The hydrodynamic helicity
has the following definition
\begin{equation}\label{TCinMF }
h=\int (\mathcal{A}\cdot \mbox{\boldmath $\Omega$}) d^{3}r
. \end{equation}
Essential part is the contraction of the elements of the vorticity.
All of them are discussed in the Introduction section.
In the previous section we show the structure of sources for the quantum vorticity.
In order to find the conserving helicity we need to demonstrate that the classical vorticity has sources of the same form,
so they can cancel each other in the equation for the evolution of the full vorticity.
Therefore, we analyse the momentum balance equation below in this section.

\subsection{On the possibility of the vector potential for the spin vorticity}

If there is the vector potential for the spin vorticity
$\mbox{\boldmath $\Omega$}_{q}=\nabla\times\mathcal{A}_{q}$
we get zero density of the topological charge $\varrho_{T}=0$.
This condition gives the possibility to find the vector potential for the spin vorticity \cite{Mahajan PRL 11}.


\subsection{The Euler equation for the velocity field}

The Euler equation for the velocity field $\textbf{v}$ leads to the classic vorticity as the
$\mbox{\boldmath $\Omega$}_{c}=\nabla\times \textbf{v}$.
Construction of the full vorticity and complete hydrodynamic helicity is considered in Ref. \cite{Mahajan PRL 11}.
Here we focus on the force fields existing for the magnetically ordered materials.

\subsubsection{Dipole-dipole interaction of the magnetic moments and the quantum Bohm potential}

We start this set of interactions with the magnetic dipole-dipole interaction and the quantum Bohm potential,
following Ref. \cite{Mahajan PRL 11} and corresponding spin torque considered above.
Corresponding force field has the following structure \cite{MaksimovTMP 2001}
\begin{equation}\label{TCinMF force DDI and qBp}
\textbf{F}=
\gamma S^{\beta}\nabla \widehat{B}^{\beta},\end{equation}
with $\widehat{B}^{\beta}$ of form of
(\ref{TCinMF magn field eff in vort}).
It leads to the equation for the classical vorticity evolution
\begin{equation}\label{TCinMF vort evol Cl with DDI}
\partial_{t}\mbox{\boldmath $\Omega$}_{c}=
\nabla\times(\textbf{v}\times\mbox{\boldmath $\Omega$}_{c})
+\gamma\nabla n^{\beta}\times\nabla \widehat{B}^{\beta}, \end{equation}
where the last term is related to spin effects and it is completely repeats
the vorticity source in equation (\ref{TCinMF vort evol Q with DDI}).

It shows that the complete vorticity,
which has no spin related sources,
is the difference between classic and quantum vorticities
$\mbox{\boldmath $\Omega$}_{\Sigma}=\mbox{\boldmath $\Omega$}_{c}-\mbox{\boldmath $\Omega$}_{q}$
\cite{Mahajan PRL 11}.
Similarly, we can construct full hydrodynamic helicity
$h_{\Sigma}=\int \mathcal{A}_{\Sigma}\cdot \mbox{\boldmath $\Omega$}_{\Sigma}d^{3}r$,
which is an integral of motion in the considered regime.

\subsubsection{Anisotropy energy associated force}

Many-particle quantum hydrodynamic method \cite{MaksimovTMP 2001}
allows to obtain the field form of the spin density evolution equation
\cite{AndreevTrukh JETP 24}, \cite{AndreevTrukh PS 24}, \cite{AndreevTrukh EPJ B 24},
discussed above,
and other hydrodynamic equations including the Euler equation,
from the quantum microscopic Pauli equation with corresponding Hamiltonian describing the interparticle interaction.
Consideration of the anisotropy energy leads to the force field
\begin{equation}\label{MFMUEI p evolution long H}
\textbf{F}=
g_{0\kappa} S^{z}\nabla S^{z}
,\end{equation}
which can be represented in form of
(\ref{TCinMF force DDI and qBp}) with
the effective magnetic field
$\textbf{B}_{eff,an}=g_{0\kappa}S^{z}\textbf{e}_{z}/\gamma$
corresponding to one considered above in the spin evolution equation.

\subsubsection{Symmetric Heisenberg exchange interaction}

Analysis of the symmetric Heisenberg exchange interaction require derivation of two nonzero terms
\begin{equation}\label{TCinMF F from HHI}
\textbf{F}=
g_{0u} S^{\beta}\nabla S^{\beta}
+\frac{1}{6}g_{u} S^{\beta}\nabla \triangle S^{\beta},\end{equation}
where the second term corresponds to the spin torque,
while the first term also appears at the derivation of the spin evolution equation,
but gives the zero contribution in the final equation.

The force field (\ref{TCinMF F from HHI}) leads to
the effective magnetic field
$\textbf{B}_{eff,ex}=(g_{0u}\textbf{S}+\frac{1}{6}g_{u}\triangle\textbf{S})/\gamma$.
The second term contributes in the cancelation of the full vorticity source
$\mbox{\boldmath $\Omega$}_{\Sigma}=\mbox{\boldmath $\Omega$}_{c}-\mbox{\boldmath $\Omega$}_{q}$
and conservation of the helicity
$h_{\Sigma}=\int \mathcal{A}_{\Sigma}\cdot \mbox{\boldmath $\Omega$}_{\Sigma}d^{3}r$.
The first term in the effective magnetic field can give the zero contribution in the full vorticity source if the concentration is constant,
similarly to the anisotropy terms.

\subsubsection{Dzylaoshinskii-Moriya interaction of helicoid type}


The force field appearing from the Dzylaoshinskii-Moriya interaction of helicoid type is
\begin{equation}\label{TCinMF F by DM 1 gamma}
\textbf{F}_{DM}=-\frac{1}{3}g_{(\gamma)}\varepsilon^{\beta\mu\nu}(S^{\beta}\nabla\partial^{\mu}S^{\nu}).
\end{equation}
It corresponds to
the effective magnetic field
$\textbf{B}_{eff,DMIh}=-\frac{1}{3}g_{(\gamma)} curl\textbf{S}/\gamma$
obtained from the spin torque considered above.
It gives the source of the classical vorticity of the same form as the source of the spin vorticity found above.
So, they can cancel each other at the consideration of the full vorticity.

\subsubsection{Dzylaoshinskii-Moriya interaction of cycloid type}

We have same picture for the Dzylaoshinskii-Moriya interaction of cycloid type as described in previous subsubsection.
We obtain the force field from the microscopic Hamiltonian of the Dzylaoshinskii-Moriya interaction of cycloid type
$$\textbf{F}_{DM}=\frac{1}{3}g_{(\beta)}
\biggl((\mbox{\boldmath $\delta$}\cdot\textbf{S})\nabla(\nabla\cdot \textbf{S})
-(\textbf{S}\cdot\nabla)\nabla(\mbox{\boldmath $\delta$}\cdot\textbf{S})\biggr)$$
\begin{equation}\label{TCinMF F by DM 1 beta}
=\frac{1}{3}g_{(\beta)}
S^{\beta}\nabla
\biggl((\delta^{\beta}(\nabla\cdot \textbf{S}))
-(\partial^{\beta}(\mbox{\boldmath $\delta$}\cdot\textbf{S}))\biggr),
\end{equation}
with
$g_{(\beta)}=\int \xi^{2}\beta(\xi)d\mbox{\boldmath $\xi$}$.
It allows to introduce
the effective magnetic field
$\textbf{B}_{eff,DMIc}=\frac{1}{3}g_{(\beta)}
(\delta^{\beta}(\nabla\cdot \textbf{S})
-\partial^{\beta}(\mbox{\boldmath $\delta$}\cdot\textbf{S}))/\gamma$.
It corresponds to the result found from the spin torque.
It also gives the source of the classical vorticity of the same form as the source of the spin vorticity found above.

\subsubsection{Dzylaoshinskii-Moriya interaction of weak ferromagnetic type for AFM}

Here we present the force field for the
Dzylaoshinskii-Moriya interaction of weak ferromagnetic type for AFM acting on the subspecies $A$ from the subspecies $B$:
\begin{equation}\label{TCinMF F DMI weakFM}
\textbf{F}_{DM,wf,A}=-g_{(0\zeta_{1})}\varepsilon^{\alpha\beta\gamma}
\delta_{2,AB}^{\alpha}S_{A}^{\beta}\nabla S_{B}^{\gamma}
.\end{equation}
Comparison with the structure of the force field of the dipole-dipole interaction (\ref{TCinMF force DDI and qBp})
shows the following form of
the effective magnetic field
$\textbf{B}_{eff,DMI,wf}=g_{(0\zeta_{1})}[\mbox{\boldmath $\delta$}_{2,AB}\times\textbf{S}_{B}]$.
It also corresponds to the result found from the spin torque
and gives the source of the classical vorticity of the same form as the source of the spin vorticity.

\subsubsection{The symmetric
exchange interaction with the odd anisotropy}

Above we present the spin torque of the OASEI and its contribution to the spin vorticity.
Here we present corresponding force fields:
\begin{equation}\label{TCinMF F OASEI A}
\textbf{F}_{OASEI,A}
=-\frac{1}{3}g_{2l_{eff}}S_{A}^{\beta}(\mbox{\boldmath $\delta$}_{eff}\cdot\nabla)\nabla S_{B}^{\beta}.
\end{equation}
Similarly, we derive the force field for the second subspecies $B$:
\begin{equation}\label{TCinMF F OASEI B}
\textbf{F}_{OASEI,B}
=\frac{1}{3}
g_{2l_{eff}}S_{B}^{\beta}(\mbox{\boldmath $\delta$}_{eff}\cdot\nabla)\nabla S_{A}^{\beta}.
\end{equation}

In accordance with equations (\ref{TCinMF force DDI and qBp}) and (\ref{TCinMF F OASEI A}),
the effective magnetic field acting on subspecies $A$ is
$\textbf{B}_{eff,OASEI,A}=-\frac{1}{3}g_{2l_{eff}}(\mbox{\boldmath $\delta$}_{eff}\cdot\nabla) \textbf{S}_{B}$.

Here we obtain equal expressions for the sources of the quantum and classical vorticities,
so they cancel each other if we construct the full vorticity.

\subsubsection{Spin-orbit interaction}

The electric dipole-dipole interaction is essential for the multiferroics,
we consider it here along with the spin-orbit interaction.
This combination leads to the spin-current model for the electric polarization of spin origin,
where the electric dipole moments are formed or oriented due to spin effects
instead of application of the external electric field.

The force field for described interactions has the following form
(derivation of the spin-orbit interaction can be found in Ref. \cite{AndreevTrukhanova 1603})
\begin{equation}\label{MFMUEI p evolution SOI eDDI}
\textbf{F}=
P^{\beta}\nabla E^{\beta}
+\frac{\gamma}{c}\varepsilon^{\beta\gamma\delta}J^{\delta\gamma}(\nabla E^{\beta})
+\frac{\gamma}{c}\partial_{t}[\textbf{E}\times \textbf{S}]
.\end{equation}
First two terms on the right-hand side of equation (\ref{MFMUEI p evolution SOI eDDI}) can compensate each other for the arbitrary inhomogeneous external or inner electric field.
In the static regime this compensation gives the balance of forces.
In the dynamical regime this compensation can be partial, so it can lead to the evolution of deformation $\textbf{u}$:
$mn\ddot{\textbf{u}}=\textbf{F}$.
It leads to the well-known spin-current model of the electric polarization appearance due to the spin effects or the magnetization
\cite{Katsura PRL 05}, \cite{Tokura RPP 14}, \cite{AndreevTrukh PS 24}
(see also discussions of the spin-current model in Refs. \cite{Andreev 2025 11} and \cite{Andreev 2025 12}).

If the static regime is formed the last term in equation (\ref{MFMUEI p evolution SOI eDDI}) is equal to zero,
but it should be compensated in order to get the static state,
so it can be combined with $mn\ddot{\textbf{u}}$.
It leads to
$\partial_{t}(mn\dot{\textbf{u}}-\frac{\gamma}{c}[\textbf{E}\times \textbf{S}])$,
under assumption of constant concentration $n=const$.
For the static spin density or its slow variation we can write
$\frac{\gamma}{c}[\textbf{E}\times \textbf{S}]\approx \partial_{t}(-\frac{\gamma}{c^{2}}[\textbf{A}\times \textbf{S}])$,
so we find
$\partial_{t}^{2}(mn\textbf{u}-\frac{\gamma}{c^{2}}[\textbf{A}\times \textbf{S}])$.
The difference in brackets can have simple, but unrealistic dependence on time,
so we assume it to be equal to zero
\begin{equation}\label{TCinMF deform by A and S}
\textbf{P}=q_{eff}\textbf{u}=\frac{\gamma}{mc^{2}}\frac{q_{eff}}{n}[\textbf{A}\times \textbf{S}]. \end{equation}
It provides a spin related deformation $\textbf{u}$,
which can be approximately associated with the ligand shift,
hence the effective charge $q_{eff}$ can be considered as the charge of the ligand.

In contrast to the spin-current model,
where we get purely spin related mechanism of the electric polarization formation,
which also allows some freedom in the chose of the spin currents of different nature and find different structures of the polarization,
in equation (\ref{TCinMF deform by A and S}) we find partially spin caused polarization and partially electromagnetic field formed polarization.
We can call it spin-field model of the electric polarization to distinguish it from the spin-current model,
while both of them appear from the spin-orbit interaction.

The last term in equation (\ref{MFMUEI p evolution SOI eDDI}) can be considered as a part of the classic vorticity $\nabla\times \textbf{v}_{eff}$
with
$\textbf{v}_{eff}=\textbf{v}-\frac{\gamma}{mnc}[\textbf{E}\times \textbf{S}]$.
However, it would require additional term with structure of the second term in equation (\ref{TCinMF vort evol Cl with DDI}).

\section{Spin-current model of the electric polarization in muliferroics}

In the Introduction to this paper we discuss general features of the spin-current model \cite{Katsura PRL 05}.
Here, we show major consequences of the generalized spin-current model related to the spin currents of different nature
\cite{AndreevTrukh JETP 24}, \cite{AndreevTrukh PS 24}.

First, let us present main formal statement of the spin-current model
\begin{equation}\label{TCinMF Spin-current model}
P^{\mu}
=\frac{\gamma}{c}\varepsilon^{\mu\alpha\beta}J^{\alpha\beta}, \end{equation}
which demonstrates that the antisymmetric spin currents forms the electric polarization.
This statement is relevant for the static regimes along with the dynamical cases.
The effective spin currents existing in the static regimes are related to the nonuniform distribution of the spin density directions.
Hence, the nonuniform distribution of the spin density or magnetization creates some distribution (or some constant value) of the electric polarization in terms of this mechanism.


\subsection{Polarization associated with symmetric Heisenberg exchange interaction}

The symmetric Heisenberg exchange interaction gives the well-known spin torque
$\textbf{T}=A[\textbf{S}\times \triangle\textbf{S}]$
which can be considered as the divergence of the spin current.
This spin current can be placed in the spin-current model of the polarization (\ref{TCinMF Spin-current model}).
It gives the macroscopic polarization
\begin{equation}\label{TCinMF P def expanded noncoll}
\textbf{P}(\textbf{r},t)=
\frac{1}{3}g_{(\alpha)}
[(\textbf{S}\cdot\nabla)\textbf{S}-\textbf{S}(\nabla\cdot \textbf{S})], \end{equation}
where
$g_{(\alpha)}=\int \alpha(r)r^{2}d^{3}r$
and function $\alpha(r)$ is proportional to the exchange integral $U(r)$: $g_{(\alpha)}\sim g_{(u)}\sim A$.
Since we get the macroscopic polarization we need to consider the possible microscopic operator of the electric dipole moment
which gives obtained polarization (\ref{TCinMF P def expanded noncoll}).
It can be shown \cite{AndreevTrukh JETP 24}, \cite{AndreevTrukh PS 24} that
the electric dipole moment of form of
\begin{equation}\label{TCinMF edm operator NONsimm}
\hat{\textbf{d}}_{ij}= \alpha_{ij}
[\textbf{r}_{ij}\times[\hat{\textbf{s}}_{i}\times\hat{\textbf{s}}_{j}]]
\end{equation}
leads to required polarization (\ref{TCinMF P def expanded noncoll}),
where $\alpha_{ij}$ is the coefficient depending on the module of relative position $\mid\textbf{r}_{ij}\mid=r_{ij}$ of spins.

Additionally, the discussion of the polarization associated with symmetric Heisenberg exchange interaction
can be found in Refs. \cite{Andreev 2025 09}, \cite{Andreev 2025 10}.

\subsubsection{Contribution of the quantum Bohm potential}

The quantum Bohm potential gives the contribution in the complete kinetic spin current tensor.
This contribution can be included in terms of the spin-current model (\ref{TCinMF Spin-current model}).
The spin current itself has the following structure
\begin{equation}\label{TCinMF spin current Bohm}
J^{\alpha\beta}_{Bohm}=-\frac{1}{2m}
\varepsilon^{\alpha\mu\nu}S^{\mu}\partial^{\beta}\biggl(\frac{S^{\nu}}{n}\biggr). \end{equation}
Therefore, it leads to polarization
$P^{\mu}_{Bohm}
=\frac{\gamma}{c}\varepsilon^{\mu\alpha\beta}J^{\alpha\beta}_{Bohm}$
which can be written in the vector notations:
\begin{equation}\label{TCinMF P Bohm fin}
\textbf{P}_{Bohm}=\frac{\gamma}{2mc}
\biggl[\textbf{S}(\nabla\cdot(\textbf{S}/n))-(\textbf{S}\cdot\nabla)\biggl(\frac{\textbf{S}}{n}\biggr)\biggr],\end{equation}
which highly resembles
polarization (\ref{TCinMF P def expanded noncoll})
obtained due to the symmetric Heisenberg exchange interaction.

If we get no concentration variation in space,
particularly the regime with no phonons excited in the system,
equations
(\ref{TCinMF P def expanded noncoll})
and
(\ref{TCinMF P Bohm fin})
show same structure.
Difference is in the coefficients,
which shows different nature of this structure.

However, the phonon propagation provides difference between equations
(\ref{TCinMF P def expanded noncoll})
and
(\ref{TCinMF P Bohm fin})
and their contribution in the dynamical properties of the multiferroics,
in scenarios similar to one considered in Ref. \cite{Risinggard SR 16}.


\subsection{Polarization associated with DMI}

\subsubsection{Polarization associated with DMI of cycloid type}

Spin torque associated with DMI of cycloid type (\ref{TCinMF s evolution for n DMI cycloid})
can be partially presented as the divergence of the effective spin current \cite{AndreevTrukh JETP 24}, \cite{AndreevTrukh PS 24}
\begin{equation}\label{TCinMF }
J^{\alpha\beta}_{DM}=-\frac{1}{6}g_{(\beta)}\varepsilon^{\alpha\beta\gamma}\delta^{\gamma}(\textbf{S}\cdot\textbf{S})
. \end{equation}
This Dzylaoshinskii-Moriya spin current can be substituted in the spin-current model (\ref{TCinMF Spin-current model}),
so it leads to the corresponding polarization $P^{\mu}_{DM}
=\frac{\gamma}{c}\varepsilon^{\mu\alpha\beta}J^{\alpha\beta}_{DM}$:
\begin{equation}\label{TCinMF P DM fin}
\textbf{P}
=-\frac{1}{3}\frac{\gamma}{c}g_{(\beta)}
\mbox{\boldmath $\delta$}(\textbf{S}\cdot\textbf{S}).
\end{equation}
Let us repeat that $\mbox{\boldmath $\delta$}$ is the ligand shift from the center of mass of the positive charge
of neighboring magnetic ions.
Basically, we can represent the polarization $\textbf{P}$ as the effective charge $q_{eff}$ multiplied on the distance between centers of masses of the positive and negative charges $\mbox{\boldmath $\delta$}$ \emph{and} the particle number density $n$:
$\textbf{P}
=q_{eff}\cdot\mbox{\boldmath $\delta$}\cdot n$.
Hence, equation (\ref{TCinMF P DM fin}) gives the effective charge density
\begin{equation}\label{TCinMF P DM fin via q}
q_{eff} n
=-\frac{1}{3}\frac{\gamma}{c}g_{(\beta)}
(\textbf{S}\cdot\textbf{S}).
\end{equation}


It can be shown that the polarization
can be found as the quantum average of the electric dipole moment
\cite{AndreevTrukh JETP 24}, \cite{AndreevTrukh PS 24}
\begin{equation}\label{TCinMF edm operator simm}
\hat{\textbf{d}}_{ij}= \mbox{\boldmath $\Pi$}_{ij} (\hat{\textbf{s}}_{i}\cdot\hat{\textbf{s}}_{j}).
\end{equation}
Here the vector constant $\mbox{\boldmath $\Pi$}_{ij}$ is used like in review \cite{Tokura RPP 14},
however comparison with the polarization (\ref{TCinMF P DM fin}) shows
$\mbox{\boldmath $\Pi$}_{ij}\sim \mbox{\boldmath $\delta$}$.

The spin torque (\ref{TCinMF s evolution for n DMI cycloid}) is an approximate macroscopic expression.
If we consider the Dzylaoshinskii-Moriya in more details
we obtain the additional spin torque containing higher space derivatives
(see eq. 44 in Ref. \cite{AndreevTrukh PS 24}).
It also allows to find additional spin current.
Hence, the spin-current model leads to the additional polarization
\begin{equation}\label{TCinMF P appr Symm}
\textbf{P}(\textbf{r},t)= \mbox{\boldmath $\delta$}_{1}
[ c_{0}(\textbf{S}\cdot\textbf{S})+c_{2}(\textbf{S}\cdot\triangle\textbf{S})]
, \end{equation}
which has same nature,
but includes the dependence on the inhomogeneity of the spin distribution.

In this regime we consider the Dzylaoshinskii constant in the form
$\textbf{D}_{ij}=\beta(r_{ij})[\textbf{r}_{ij}\times\mbox{\boldmath $\delta$}_{1}]$.
So, we can present relations between parameters
$\beta(r_{ij})$, $c_{0}$, $c_{2}$ and $\mbox{\boldmath $\Pi$}_{ij}(r_{ij})$ \cite{AndreevTrukh PS 24}:
$c_{0}\mbox{\boldmath $\delta$}_{1}=\int \mbox{\boldmath $\Pi$}_{ij}(r_{ij})d^{3}r_{ij}$,
$c_{2}\mbox{\boldmath $\delta$}_{1}=(1/6)\int r_{ij}^{2}\mbox{\boldmath $\Pi$}_{ij}(r_{ij})d^{3}r_{ij}$,
and
$\frac{\partial \mbox{\boldmath $\Pi$}}{\partial r}
=\frac{\gamma}{c}r\beta(r)\mbox{\boldmath $\delta$}_{1}$.
The last equation can be considered in different form:
$\mbox{\boldmath $\Pi$}
=-\frac{\gamma}{3c}r^2\beta(r)\mbox{\boldmath $\delta$}_{1}$.

This result can be generalized on the antiferromagnetic multiferroics \cite{AndreevTrukh PS 24}
\begin{equation}\label{TCinMF P def expanded AFR}
\textbf{P}=\mbox{\boldmath $\delta$}
[2 c_{0,AB}
(\textbf{S}_{A}\cdot\textbf{S}_{B})
+c_{2,AB}
(S^{\nu}_{A} \triangle S^{\nu}_{B} +S^{\nu}_{B} \triangle S^{\nu}_{A})].
\end{equation}

Additionally, the discussion of the polarization associated with the DMI of cycloid type
can be found in Refs. \cite{Andreev 2025 09}, \cite{Andreev 2025 10}.

\subsubsection{Polarization associated with DMI of helicoid type}

Let us to point out that
there is no
electric polarization of spin origin associated with the DMI of helicoid type,
which appears via the spin-current mechanism,
since the DMI of helicoid type leads to the symmetric spin current.

\subsection{Polarization associated with OASEI}

We presented above the contribution of the odd anisotropy
of the symmetric
exchange interaction (OASEI)
in the spin torque and the force field.
We also demonstrated corresponding structure of the exchange integral (\ref{TCinMF U structure in Ham HHI 2}).
In this case the spin torque allows to find the effective spin current
as the composition of the partial spin currents in each subsystem of the AFM \cite{Andreev 2025 12}
$J_{\Sigma}^{\alpha\beta}=J_{A}^{\alpha\beta}+J_{B}^{\alpha\beta}
=\frac{1}{3}\varepsilon^{\alpha\mu\nu}g_{2l_{eff}}\delta_{eff}^{\beta}S_{A}^{\mu}S_{B}^{\nu}$.

So, we obtain another mechanism of the electric polarization formation,
which exists in multicomponent magnetically order structures,
such as the AFM multiferroics \cite{Andreev 2025 12}:
\begin{equation}\label{TCinMF Pol nm OASEI}
\textbf{P}=\frac{1}{3}\frac{\gamma}{c}g_{(l)}
[\textbf{S}_{B}(\textbf{S}_{A}\cdot\mbox{\boldmath $\delta$}_{eff})
-\textbf{S}_{A}(\textbf{S}_{B}\cdot\mbox{\boldmath $\delta$}_{eff})].
\end{equation}
Polarization (\ref{TCinMF Pol nm OASEI}) show some similarity to
polarization
(\ref{TCinMF P def expanded noncoll}),
but instead of the derivatives we get the ligand shifts.
This configuration provides
the tensor effective charge
$P^{\alpha}=q_{eff}^{\alpha\beta}n\delta_{eff}^{\beta}$
with
$$q_{eff}^{\alpha\beta}=\frac{1}{3n}\frac{\gamma}{c}g_{(l)}
[S_{B}^{\alpha}S_{A}^{\beta}
-S_{A}^{\alpha}S_{B}^{\beta}]$$
\begin{equation}\label{TCinMF }
=\frac{1}{6n}\frac{\gamma}{c}g_{(l)}
[M^{\alpha}L^{\beta}
-L^{\alpha}M^{\beta}],
\end{equation}
where the charge is proportional to the antisymmetric structure of partial spin densities
or the antisymmetric combination of the antiferromagnetic vectors
$\textbf{L}=\textbf{S}_{A}-\textbf{S}_{B}$, and $\textbf{M}=\textbf{S}_{A}+\textbf{S}_{B}$.

Polarization (\ref{TCinMF Pol nm OASEI}) can be found from the following microscopic electric dipole moment:
\begin{equation}\label{TCinMF}
\hat{\textbf{d}}_{ij}=-\frac{1}{6}\frac{\gamma}{c}r_{ij}^{2}l(r_{ij})
[\mbox{\boldmath $\delta$}_{eff,ij}\times [\hat{\textbf{S}}_{i}\times \hat{\textbf{S}}_{j}]].
\end{equation}

\section{Reverse spin-current model: spin-current formed by the electric polarization}

The spin evolution equation contains the spin-orbit interaction,
which is proportional to the spin-current.
Moreover, it contains the antisymmetric part of the spin-current tensor.
Therefore, we expect that a part of the spin current can be associated with the electric polarization
(other part can exist in the fluid systems)
in accordance with the spin-current model
(\ref{TCinMF Spin-current model}).

Formation of the effective spin current related to the current of magnons or some static nonuniform spin distribution
is called here as the reverse SCM.
Let us additional state that the polarization can form the antisymmetric spin current only.
But corresponding spin distribution can be considered as the mechanism of the spin polarization or spin noncollinear structure
(with some partial polarization formation due to the electric polarization or the deformation in ionic crystals).

The reverse spin-current model states that
the polarization created in the system creates the antisymmetric effective spin current
\begin{equation}\label{TCinMF }
\frac{\gamma}{c}J^{\alpha\beta}=\frac{1}{2}\varepsilon^{\alpha\beta\gamma}P^{\gamma}
, \end{equation}
where the effective spin current can be related to some nonuniform static distribution of the spin directions.
Hence, this model shows formation of the noncollinear order of spins if spins are in the ferromagnetic state.
However, we can consider it in more general way.
It can be considered as the mechanism of formation of the spin polarization in the system of unpolarized atoms/ions in order to create the spin current which requires some partial spin polarization.

If the spin-current model cancels the spin-orbit interaction contribution in the Euler equation giving the deformation dynamics via the electric dipole-dipole interaction,
we have no similar relation in the spin evolution equation,
which nevertheless contains a term caused with the spin-orbit interaction.
But the reverse spin-current model gives the cancelation of this term.

The spin orbit torque in the spin evolution equation
\begin{equation}\label{TCinMF }
T^{\alpha}_{SO}=-\frac{\gamma}{ c}\varepsilon^{\alpha\beta\gamma}\varepsilon^{\beta\mu\nu}E^{\mu} J^{\gamma\nu}
, \end{equation}
can be transformed into
\begin{equation}\label{TCinMF }
\textbf{T}_{SO}=\frac{1}{2}
[\textbf{E}\times \textbf{P}]
. \end{equation}

Hence, we obtain zero contribution of the SOI if the spin current is caused with the electric polarization,
which is cased by the electric field $\textbf{P}(\textbf{E})=\kappa \textbf{E}$.
This conclusion is less useful if the electric field is not parallel to the main axes of the crystal,
so the tensor nature of this relation is applied $P^{\alpha}=\kappa^{\alpha\beta}E^{\beta}$.

If we obtain relation $\textbf{P}(\textbf{E})=\kappa \textbf{E}$ in the static limit
(as the response on the external field),
we can get some variation of the polarization due to the dynamic of the system,
so the polarization can influence the spin dynamics via the spin-orbit interaction.

The physical mechanisms for the formation of the magnetization due to the electric polarization are discussed in Ref. \cite{JuraschekBalatsky PRM 17}.
The qualitative analysis of the motion of charges in ion crystals leading to the magnetization (circle currents) formation is given.
In this section we consider this problem in the formal way with reference to the spin-current model and structure of the hydrodynamic equations.
However, it allows us to conclude that the polarization can cause the spin current rather then the magnetization itself.
The effective spin currents, such as the Heisenberg or Dzylaoshinskii-Moriya spin currents,
exist at the static nonuniform distribution of the spins (particularly their directions) in space.
So, the formation of the spin current requires the formation of the magnetization as well.

\subsection{Role of the reverse spin-current model in the helicity conservation}

The spin-current model (\ref{TCinMF Spin-current model}) allows to cancel the force fields
which gives the sources of the classical vorticity
(like the electric dipole-dipole interaction force).
The concept of the spin-field model of the electric polarization suggested in this paper allows to cancel the second part of the spin-orbit interaction,
which can be a source of the classical vorticity and the full helicity as well.
Cancelation of the contribution of the spin-orbit interaction in the Euler equation gives two mechanisms for the electric polarization formation of the spin or field-spin origin.
The reverse spin-current model in the spin evolution equation allows to cancel the contribution of the spin-orbit interaction spin torque,
which can form the source of the quantum vorticity.
However, physical consequence of the reverse spin-current model is in mechanism of the spin current formation
(and corresponding nonuniform magnetization distribution)
due to the electric polarization


\section{Spin-field model of the electric polarization}

Here we repeat equation (\ref{TCinMF deform by A and S}) found at the analysis of the balance of forces created via the spin-orbit interaction:
\begin{equation}\label{TCinMF deform by A and S 2}
\textbf{u}=\frac{\gamma}{mc^{2}}\frac{1}{n}[\textbf{A}\times \textbf{S}]. \end{equation}
It provides a spin related deformation $\textbf{u}$,
which appears from the spin-orbit interaction force field in the Euler equation.
In contrast with the spin-current model,
which contains the spin current tensor appearing from additional sources and showing different nature,
equation (\ref{TCinMF deform by A and S 2}) presenting the spin-field model of the electric polarization directly refers to
the spin density $\textbf{S}$ and characteristic of the electromagnetic field $\textbf{A}$.
The vortical part of the electric field is parallel to the vector potential $\textbf{A}$,
hence the deformation is perpendicular to the plane formed by the electric field and the spin density.
This is the partial deformation in the system existing due to the spin effects in the presence of the electric field.

In this paper we consider the models of the magnetoelectric effect from the point of view of interactions between ions/atoms.
Generalization of the spin-current model on the different types of the electric polarization of spin origin is the central part of the presented analysis,
while other methods of the polarization description can be found in recent literature \cite{Solovyev PRL 21}, \cite{Mostovoy npj 24}.
We present obtained models in the formal way, with no application to the equilibrium spin structures \cite{Gareeva PRB 13},
their stability \cite{Gareeva PRB 13}, or the spectrum of the perturbations
\cite{Andreev 2025 09}, \cite{Andreev 2025 10},
\cite{Andreev 2025 05}, \cite{Castro PRB 25}, \cite{AndreevTrukh EPL 25},
\cite{Fishman PRB 19}, \cite{Holbein PRB 23}.
We do not discuss the methods of the microscopic derivation of the presented equations
since they are described in earlier papers \cite{MaksimovTMP 2001}, \cite{Andreev 2025 Vestn},
or standard method can be used as well \cite{KOSEVICH PR 90}.

\section{Conclusion}

The spin-current model of the electric polarization has been considered as a mechanism for the hydrodynamic helicity conservation
appearing via the cancelation of two force fields (the electric dipole-dipole interaction and a part of the spin-orbit interaction),
which have same dependence on the nonuniform electric field.
However, in nonstatic regime, there is the second part of the spin-orbit interaction force field,
its cancelation leads to novel model of the deformation formation
(and the electric polarization formation as well)
in addition to the spin-current model.
This model has been called
the spin-field model of the electric polarization.

These models have been suggested and discussed in context of the conservation of the topological charge and the vorticity along with the hydrodynamic helicity.
The systematic analysis of the force fields in the Euler equation for the velocity field and the spin torque field in the LLG equation has demonstrated
that they contribution in the spin and classic vorticity evolution equations can cancel each other,
so no spin-related sources gives contribution in the full vorticity.
Hence, the conservation of the topological charge and the hydrodynamic helicity is shown via the dynamical model,
since both the topological charge and the hydrodynamic helicity can be presented via vorticity.
Moreover, the conservation of the topological charge gives more strong restrictions on the structure of the source of the spin vorticity,
which corresponds to the restrictions on the spin-torque.
The spin-orbit interaction gives the spin torque showing a deviation from required structure.
Therefore, it should be balanced or canceled.
The model for the approximate cancelation
(in the static regime)
has been suggested.
It has been called the reverse spin-current model
of the nonuniform magnetization formation due to the polarization in the system.

The spin-orbit interaction
plays
the crucial role in the description of the magnetoelectric effect,
which is a relatively small effect in the magnetic materials.
All models discussed here
(the spin-current model of the electric polarization,
the reverse spin-current model of the magnetization,
and the spin-field model of the deformation)
appears as the result of analysis of the spin-orbit interaction in different equations.

We do not repeat all interactions included in the analysis of the vorticity behavior.
We mention the three forms of the Dzylaoshinskii-Moriya interaction (including one existing in the AFM only),
the quantum part of the spin-current divergence,
and recently suggested ligand shift related modification of the symmetric Heisenberg exchange interaction
called the symmetric exchange interaction with the odd anisotropy.
All these terms are in agreement with the requirements for the conservation of
the topological charge and the hydrodynamic helicity.

\section{DATA AVAILABILITY}


Data sharing is not applicable to this article as no new data were
created or analyzed in this study, which is a purely theoretical one.

\section{Acknowledgements}

The work is supported by the Russian Science Foundation under the grant No. 26-79-32004.




\end{document}